\documentclass{elsart}

\usepackage[]{graphicx}
\begin{document}
\begin{frontmatter}
\title{Magnetic properties of 
the two-dimensional Heisenberg model on a triangular lattice}
\author{P. Rubin\thanksref{mail}},
\author{A. Sherman}
\address{Institute of Physics, University of Tartu,
Riia 142, 51014 Tartu, Estonia}
\thanks[mail]{Corresponding author:
E-mail: rubin@fi.tartu.ee}

\begin{abstract}

The spin Green's function of the antiferromagnetic Heisenberg
model on a triangular lattice is calculated  using Mori's
projection operator technique. At $T=0$  the spin excitation
spectrum is shown to have
 gaps at the wave vectors of the classical N\'{e}el ordering.
This points to the absence of the antiferromagnetic long-range
order in the ground state. The calculated spin correlation on the
neighboring sites of the same sublattice is in good agreement with
the value derived from exact diagonalization. The temperature
dependencies of the spin correlations and the gaps are calculated.

\end{abstract}
\end{frontmatter}

Strong electron correlations are expected to play an essential
role in the recently discovered electron doped cobaltate
 Na$_x$CoO$_2$*$y$H$_2$O \cite{Takada,Li,Yang,Motrunich}. This system
has layered structure
which 
consists 
of two-dimensional CoO$_2$ planes separated by Na$^+$ and H$_2$O layers
and resembles the structure of cuprate high-$T_c$ superconductors.
However, in contrast to square CuO$_2$ planes the Co ions form a
triangular lattice. It is supposed that for  moderate Na
concentrations $x$ the low-energy physics of CoO$_2$
planes  can be described by 
the $t$-$J$ model \cite{Baskaran}. For small $x$ the major part of
Co ions are in the Co$^{4+}$ state with the spin $S=1/2$ and the
$t$-$J$ model is reduced to the two-dimensional $S=1/2$ Heisenberg
model. The knowledge of the spin excitation spectrum of this model
is necessary for further investigation of spectral and magnetic
properties of the CoO$_2$ planes. Various numerical and analytical
methods were used for elucidating the ground state properties of
the Heisenberg model on
a triangular lattice 
\cite{Capriotti}.
 In Refs.~\cite{Capriotti,Shimizu} it is argued
that the ground state of this system has the long-range
antiferromagnetic order. The opposite conclusion  was made in
Refs.~\cite{Nishimori,Runge} analyzing the data of the exact
diagonalization of small clusters.

In this work we use the method \cite{Sherman} based on Mori's
projection operator technique \cite{Mori} in which the spin
Green's function is represented by a continued fraction. The
elements of the  fraction are calculated in the recursive
procedure which is similar to Lanczos' orthogonalization. The
decoupling is used in calculating the terms of the fraction. By
analogy with Ref.~\cite{Kondo} the decoupling is corrected by
introducing  vertex parameters the values of which are determined
from the constraint of zero site magnetization following from the
rotational symmetry of the Hamiltonian  and the nearest-neighbor
spin correlation derived from exact diagonalization. The used
approach retains the rotation symmetry of spin components and does
not contain any preset magnetic ordering. With this procedure the
energies of magnetic excitations and spin-spin correlations are
calculated. The calculated spin correlation on  neighboring sites
of the same sublattice is in good agreement with the value
available from exact diagonalization \cite{Nishimori,Runge}. The
spectrum of magnetic excitations is found to have finite gaps at
wave vectors of the classical N\'{e}el ordering
 $\textbf{Q}=\left( \frac{2 \pi }{3}, \frac{2 \pi }{\sqrt{3}}
\right), \left( \frac{4 \pi }{3}, 0 \right)$ for zero temperature.
This indicates that  the ground state of the Heisenberg model on a
triangular lattice has no long-range magnetic ordering.

The Hamiltonian of this  model  reads
\begin{equation}\label{hamiltonian}
H=\frac{1}{2}\sum_{\bf nm}J_{\bf nm}\left(s^z_{\bf n}s^z_{\bf
m}+s^{+1}_{\bf n}s^{-1}_{\bf m}\right),
\end{equation}
where $s^z_{\bf n}$ and $s^\sigma_{\bf n}$ are the components of the
spin-$\frac{1}{2}$ operators, {\bf n} and {\bf m} label sites of the
triangular lattice, $\sigma=\pm 1$. 
 We take
into account nearest neighbor interactions only,  $J_{\bf
nm}=J\sum_{\bf a}\delta_{\bf n,m+a}$ where $J$ is the exchange
constant and the six vectors {\bf a} connect nearest neighbor
sites.

Magnetic properties of the model are inferred from the  spin
retarded Green's function
\begin{equation}\label{green}
 D({\bf k}t)=-i\theta(t)\langle[s^z_{\bf k}(t),s^z_{\bf
-k}]\rangle,
\end{equation}
where  
$s^z_{\bf k}=N^{-1/2}\sum_{\bf n}\exp(-i{\bf kn})s^z_{\bf n}$, $N$
is the number of sites, $s^z_{\bf k}(t)=\exp(iHt)s^z_{\bf
k}\exp(-iHt)$ and $\langle\ldots\rangle={\rm
Sp}[\exp(-H/T)\ldots]/{\rm Sp} [\exp(-H/T)]$ with the
temperature $T$. 
As mentioned, we apply Mori's projection operator technique
\cite{Sherman,Mori}. To obtain the spin Green's function we use 
the continued fraction representation for Kubo's relaxation function 
$(\!( s^z_{\bf k}|s^z_{\bf -k})\!)_\omega=\int_{-\infty}^\infty
dt\exp(i\omega t)(\!( s^z_{\bf k}|s^z_{\bf -k})\!)_t$, $(\!( s^z_{\bf
k}|s^z_{\bf -k})\!)_t=\theta(t)\int_t^\infty dt'\langle[s^z_{\bf k}(t),
s^z_{\bf - k}]\rangle$,
\begin{equation}\label{cfraction}
(\!( s^z_{\bf k}|s^z_{\bf -k})\!)_\omega=\frac{\displaystyle(s^z_{\bf
k},s^z_{\bf -k})}{\displaystyle \omega-E_0-\frac{\displaystyle
V_0}{\displaystyle\omega-E_1-\frac{\displaystyle V_1}{\ddots}}},
\end{equation}
and the relation 
\begin{equation}\label{gk}
 D({\bf \omega}t)=\omega (\!( s^z_{\bf k}|s^z_{\bf
-k})\!)_\omega -(s^z_{\bf k},s^z_{\bf -k}),
\end{equation} 
where $(A,B)=i\int_0^\infty dt\langle[A(t),B]\rangle$.

The elements of the fraction $E_i$ and $V_i$ are determined from the
recursive procedure 
\begin{eqnarray}
&&[A_n,H]=E_nA_n+A_{n+1}+V_{n-1}A_{n-1},\quad E_n=([A_n,H],
 A_n^\dagger)\,(A_n,A_n^\dagger)^{-1},\nonumber\\[-0.5ex]
&&\label{lanczos}\\[-0.5ex]
&&V_{n-1}=(A_n,A_n^\dagger)\,(A_{n-1},
A_{n-1}^\dagger)^{-1},
 \quad V_{-1}=0, \quad A_0=s^z_{\bf k},\quad n=0,1,2,\ldots\nonumber
\end{eqnarray}
The operators $A_i$ constructed in this procedure
form an orthogonal set, 
$(A_i,A^\dagger_j)\propto\delta_{ij}$.

Using procedure (\ref{lanczos}) we get
\begin{eqnarray*}
&&E_0=(i\dot{s}^z_{\bf k},s^z_{\bf -k})(s^z_{\bf k},s^z_{\bf
-k})^{-1}=\langle[s^z_{\bf k},s^z_{\bf -k}]\rangle(s^z_{\bf k},s^z_{\bf
-k})^{-1}=0,\quad A_1=i\dot{s}^z_{\bf k},\\
&&F_0=6J|C_1|(1-\gamma'_{\bf k})(s^z_{\bf k},s^z_{\bf -k})^{-1},\quad
E_1=(i^2\ddot{s}^z_{\bf k},-i\dot{s}^z_{\bf -k})(i\dot{s}^z_{\bf
k},-i\dot{s}^z_{\bf -k})^{-1}=0,
\end{eqnarray*}
where $\gamma'_{\bf k}=\frac{1}{3}\cos(k_x)+\frac{2}{3}
\cos(\frac{k_x}{2}) \cos(\frac{k_y \sqrt3}{2})$ in the orthogonal
system of coordinates, $C_1 = \langle  s_{\bf n}^{+1} s_{\bf n +
a}^{-1} \rangle$ is the spin correlation  on neighboring sites. At
this point we interrupt the continued fraction and calculate
$(s^z_{\bf k},s^z_{\bf -k})$. In the accepted approximation $V_1
\propto (A_2 ,A_2^\dagger) = 0$. From this equation we find
\cite{Sherman}
\begin{equation}\label{aii}
\langle[i^2\ddot{s}^z_{\bf k},-i\dot{s}^z_{\bf -k}]\rangle=36J^2C_1^2
(1-\gamma'_{\bf k})^2(s^z_{\bf k},s^z_{\bf -k})^{-1}.
\end{equation}
The quantity $i^2\ddot{s}^z_{\bf k}$ in the left-hand side of this
equation contains terms of the type $s^z_{\bf l}s^{+1}_{\bf
n}s^{-1}_{\bf m}$ which, following Refs.~\cite{Kondo,Shimahara91},
are approximated by $[(\alpha_1 P_{\bf{nm}} + \alpha_2 (1 -
P_{\bf{nm}} ))\langle s^{+1}_{\bf n}s^{-1}_{\bf
m}\rangle(1-\delta_{\bf nm})+\frac{1}{2}\delta_{\bf nm}]s^z_{\bf
l}$. Here $\alpha_{1,2}$ are the vertex corrections, $P_{\bf{nm}}=
1$ if the sites $\bf {n}$ and $\bf {m}$ belong to the same
magnetic sublattice and zero in the opposite case. In the
decoupling the constraint of zero site magnetization
\begin{equation}\label{constraint}
\left\langle s^z_{\bf n}\right\rangle =\frac{1}{2}-\left\langle
s^{-1}_{\bf n}s^{+1}_{\bf n}\right\rangle=0
\end{equation}
was taken into account. This constraint is fulfilled both in the
paramagnetic and in the ordered states; in the latter case due to
the averaging over all possible directions of the magnetization.
Using this approximation for $i^2\ddot{s}^z_{\bf k}$ from
Eq.~(\ref{aii}) we find $(s^z_{\bf k},s^z_{\bf -k})$ and from
Eqs.~(\ref{cfraction}) and (\ref{gk}) we get
\begin{equation}\label{gf}
D({\bf k}\omega)=\frac{6(1-\gamma^\prime_{\bf
 k})J|C_1|}{\omega^2-
 \omega^2_{\bf k}},
\end{equation}
where
\begin{equation}\label{energies}
\omega^2_{\bf k}=36J^2\alpha_2 |C_1|(1-\gamma'_{\bf k})
\left(\Delta+\frac{1}{2}+\gamma'_{\bf k}\right),\quad
\Delta=\frac{\alpha_1 C_2^\prime + \alpha_2 C_2^{ \prime
\prime}}{\alpha_2 |C_1|} -\frac{1}{3} + \frac{(1-\alpha_1) }{12
\alpha_2 |C_1|},
\end{equation}
where  $ C_2^\prime = \frac{1}{2} + 2 \langle s_{\bf n}^{+1}s_{\bf
n+ a''}^{-1}\rangle $ with the vector ${\bf a''}$ connecting
nearest neighbor sites of a magnetic sublattice and $C_2^{\prime
\prime} = 2 C_1 + \langle s_{\bf n}^{+1}s_{\bf n+ 2 a}^{-1}\rangle
 $. The quantity $\omega_{\bf k}$ is the frequency of spin excitations
and the parameter $\Delta$ describes the spin gap at the wave
vectors {\bf Q} of the classical N\'eel ordering. As  seen from
Eq.~(\ref{energies}), the frequency of  spin excitations tends to
zero when $\bf k \rightarrow 0$ (the Goldstone mode) and has local
minima at $\bf k = Q$.

To finish the calculations we have to find the parameters $\alpha
_1 $, $\alpha _2 $, $C_1 $, ${C}'_2 $ and ${C}''_2 $ in Eq. (8)
and (9). For this purpose we use Eq. (7) and the relations
connecting the spin correlations with Green's function (8)
\begin{equation}
\label{eq1} \left\langle {s_{\rm {\bf n}}^+ s_{\rm {\bf m}}^- }
\right\rangle =\frac{6J\left| {C_1 } \right|}{N}\sum\limits_{\rm
{\bf k}} \cos{\left({\rm {\bf k}}\left( {{\rm {\bf n}}-{\rm {\bf
m}}} \right) \right)}
\frac{1-{\gamma }'_{\rm {\bf k}} }{\omega _{\rm {\bf k}} }\coth
\left( {\frac{\omega _{\rm {\bf k}}}{2 T}} \right).
\end{equation}
However, these relations -- the three equations (\ref{eq1}) for
$C_1 $, ${C}'_2 $, ${C}''_2 $ and the constraint (7) -- give only
four equations for the above five parameters. At zero temperature
this deficiency can be made up by using results of exact
diagonalization. More specifically, we use the value  $C_1
=-0.1215$ derived for an infinite crystal from results of exact
diagonalization [8]. If the assumption is made that for $T=0$ the
system has the long-range antiferromagnetic order then the
frequencies of spin excitations at \textbf{Q} have to vanish. In
this case one more unknown parameter -- the condensation part $C$
-- appears in the calculations [13]. This part is the contribution
of points \textbf{Q }to the spin correlations (\ref{eq1}),
\begin{equation}
\label{tzero} \left\langle {s_{\rm {\bf n}}^+ s_{\rm {\bf m}}^- }
\right\rangle =C \cos{\left({\rm {\bf Q}}\left( {{\rm {\bf
n}}-{\rm {\bf m}}} \right) \right)}+  \frac{6J\left| {C_1 }
\right|}{N}\sum\limits_{\rm {\bf k \neq Q}}
\cos{\left({\rm {\bf
k}}\left( {{\rm {\bf n}}-{\rm {\bf m}}} \right) \right)}
\frac{1-{\gamma }'_{\rm {\bf k}} }{\omega _{\rm {\bf k}} }.
\end{equation}
Also the above set of equations is supplemented by  the equation
$\Delta =0$ with $\Delta$ given by Eq. (\ref{energies}). The
sublattice magnetization $m$ is directly connected with the
condensation part: $m=\sqrt {{3C} \mathord{\left/ {\vphantom {{3C}
2}} \right. \kern-\nulldelimiterspace} 2} .$ Solving the mentioned
set of equations for an infinite crystal we found that $C=-0.05.$
This result contradicts to the assumption of the long-range order
for which $C$ has to be positive. We arrived at the conclusion
that either there is no long-range order in the ground state of
the 2D Heisenberg model on the triangular lattice or the ordering
parameter is too small to be determined by the present approximate
method. Notice also that the value of $\langle s_{\bf
n}^{+1}s_{\bf n+ a''}^{-1}\rangle = 0.077$ obtained in our
calculations is in good agreement with the value $0.0803 \pm
0.004$ derived from exact diagonalization [8] (to our knowledge
$C_1$ and $ \langle s_{\bf n}^{+1}s_{\bf n+ a''}^{-1} \rangle$ are
the only correlations with the values well-defined from exact
diagonalization for the triangular lattice). One of the
manifestations of the short-range order is a finite frequency
$\omega _{\rm {\bf Q}} $ of spin excitations at the wave vectors
\textbf{Q}.

To consider the temperature variation of the spin correlations and
spectrum we decreased the number of parameters, setting $\alpha _1
=\alpha _2 =\alpha .$ In this approximation we have four equations
for four parameters. The range of temperatures $0\le T\le 1.3 J$
and a $1000\times 1000$ lattice with periodic boundary conditions
were used in these calculations. The approximation $\alpha _1
=\alpha _2 $ is cruder than the previous one where $\alpha _1 $
and $\alpha _2 $ could differ. Therefore the gap $\omega _{\rm
{\bf Q}} $ in the spectrum of spin excitations at the wave vectors
\textbf{Q}, the existence of which is an indication of the
short-range order, is somewhat larger in the former approximation
than in the latter. The temperature dependencies of this gap, the
spin correlation $C_1$ and the dispersion of the spin excitations
 in the approximation $\alpha _1 =\alpha _2 $ are shown in Fig. 1
-- 3. The gap $\omega _{\rm {\bf Q}} $ grows with temperature
which corresponds to the decrease of the correlation length.

In summary, Mori's projection operator technique was used for
investigating the excitation spectrum of the two-dimensional
Heisenberg model on a triangular lattice. The zero-temperature
spin-excitation spectrum has  gaps at the wave vectors $\left(
{\frac{2\pi }{3},\frac{2\pi }{\sqrt 3 }} \right)$, $\left(
{\frac{4\pi }{3},0} \right)$ of the classical N\'{e}el ordering
which is an indication of the lack of the long-range
antiferromagnetic ordering in the ground state of the system.
Temperature dependencies of the spin correlations and the gaps
were calculated.

\section*{Acknowledgements}
This work was supported by the ESF Grant No 5548.

\newpage

Figure captions

Fig.~1.  The dependence of the spin correlation $ C_1 = \langle
s_{\bf n}^{+1} s_{\bf n + a}^{-1}\rangle$ on temperature.

Fig.~2.  The dependence of the energy of the spin excitations  at
the wave vector of the classical N\'{e}el ordering ${\bf Q}$ on
temperature.

Fig.~3.  The dispersion of the spin excitations
 at zero temperature.

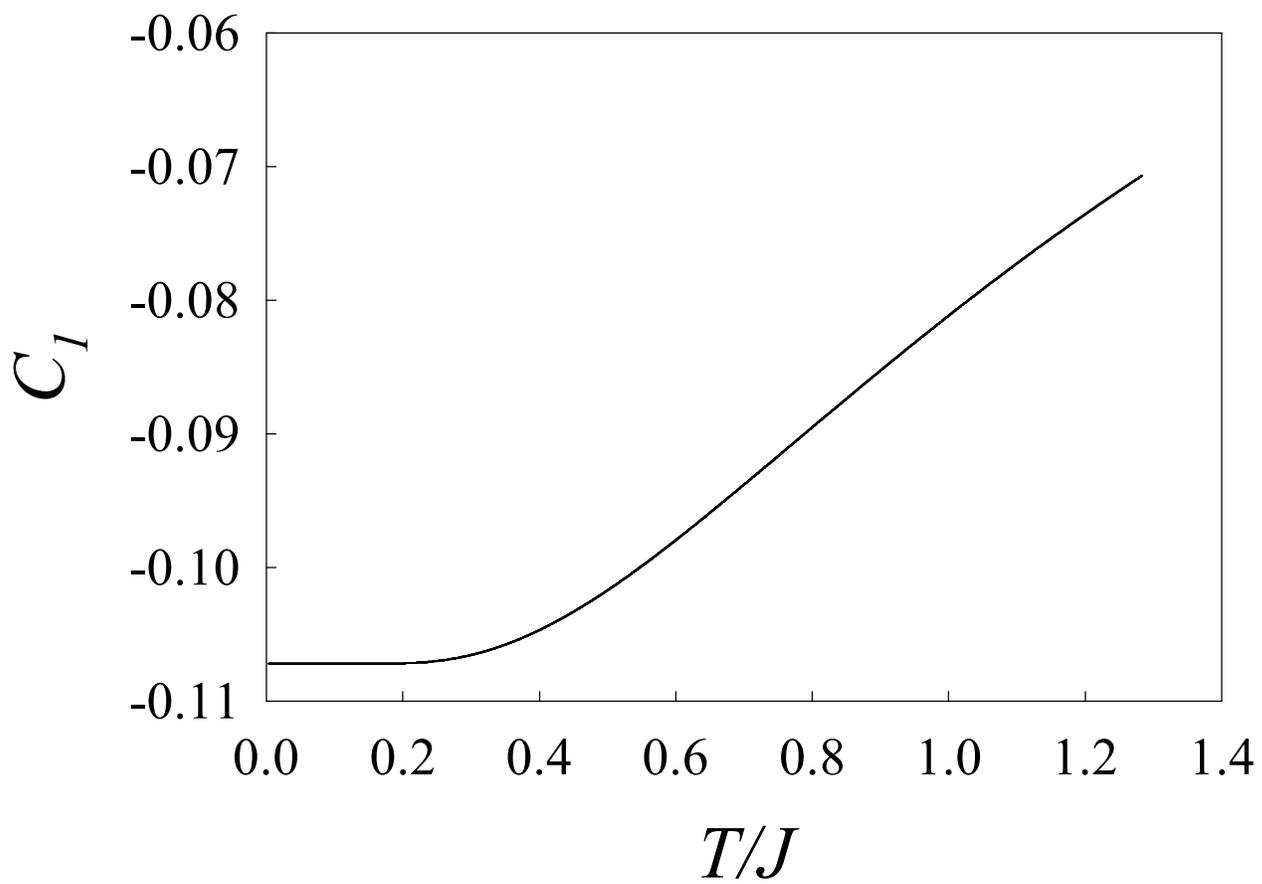
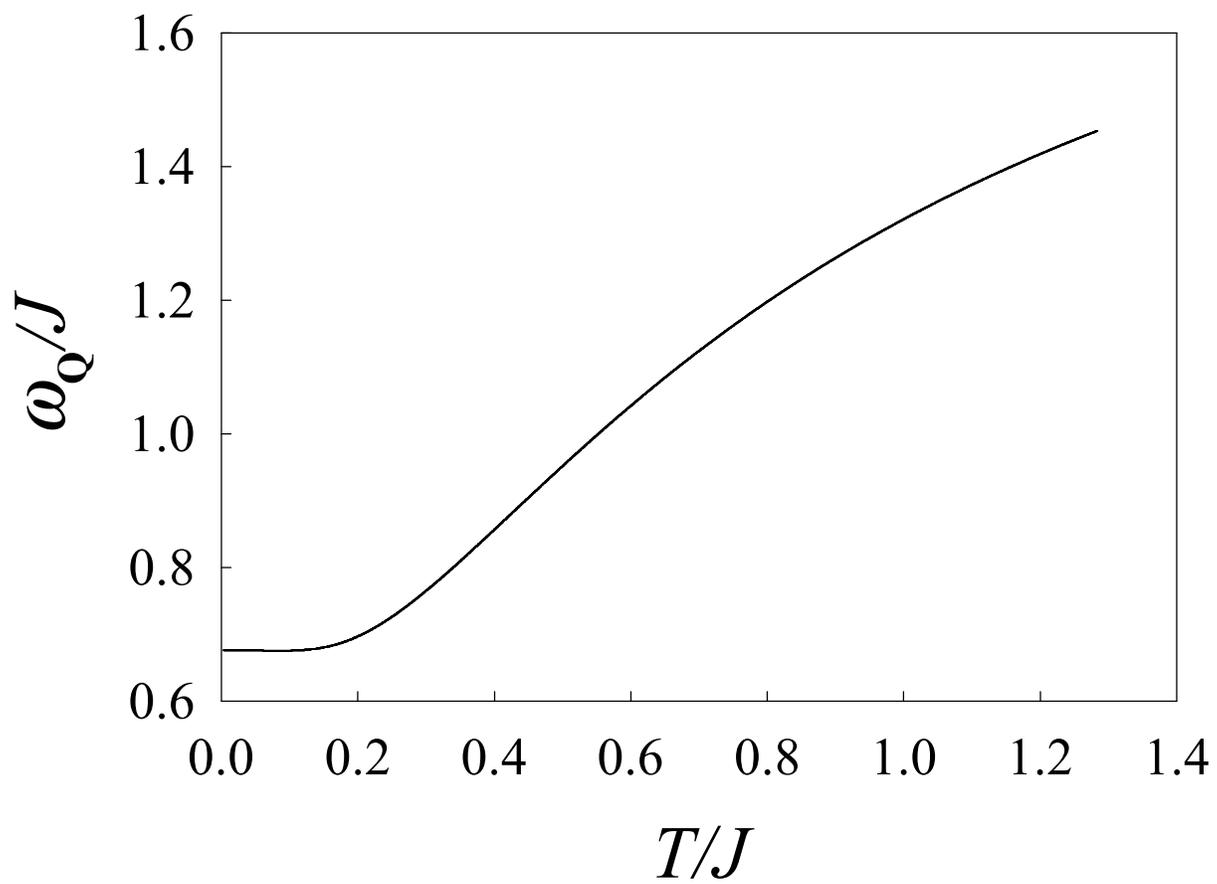
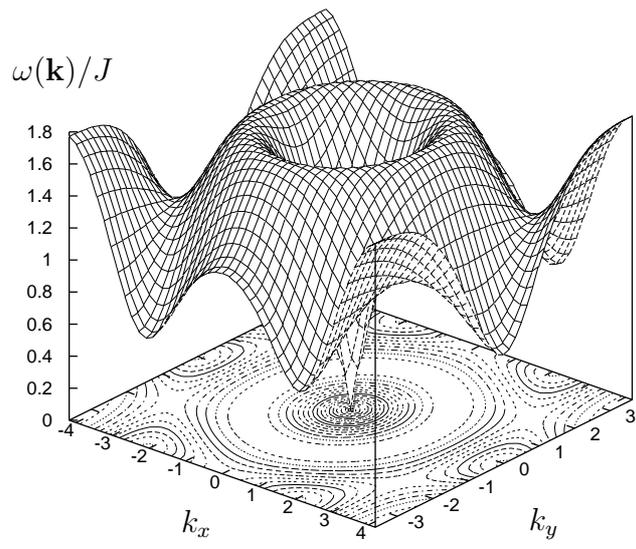
\end{document}